\documentclass[runningheads]{llncs}
\usepackage{graphicx}
\begin{document}
	\title{GQE-PRF: Generative Query Expansion with Pseudo-Relevance Feedback}
	%
	%
	\author{Minghui Huang \and Dong Wang \and Shuang Liu \and Meizhen Ding}
	%
	%
	\institute{Artificial Intelligence Application Research Center, Huawei Technologies, \\ 
		\email{\{huangminghui3, wangdong153, liushuang30, dingmeizhen\}@huawei.com}\\
	}
	\maketitle              

	\begin{abstract}
		Query expansion with pseudo-relevance feedback (PRF) is a powerful approach to enhance the effectiveness in information retrieval. Recently, with the rapid advance of deep learning techniques, neural text generation has achieved promising success in many natural language tasks. To leverage the strength of text generation for information retrieval, in this article, we propose a novel approach which effectively integrates text generation models into PRF-based query expansion. In particular, our approach generates augmented query terms via neural text generation models conditioned on both the initial query and pseudo-relevance feedback. Moreover, in order to train the generative model, we adopt the conditional generative adversarial nets (CGANs) and propose the PRF-CGAN method in which both the generator and the discriminator are conditioned on the pseudo-relevance feedback. We evaluate the performance of our approach on information retrieval tasks using two benchmark datasets. The experimental results show that our approach achieves comparable performance or outperforms traditional query expansion methods on both the retrieval and reranking tasks.	
		
	\end{abstract}
	\section{Introduction}
	Query Expansion (QE)~\cite{DBLP:conf/sigir/Voorhees94,DBLP:journals/ipm/BhogalMS07,DBLP:journals/csur/CarpinetoR12,DBLP:journals/ipm/AzadD19} is a common technique to improve the performance of information retrieval approaches. It augments extra semantically related terms to the original query and thus alleviates the lexical mismatch problem between query and document terms. Clearly, one of the key factors to the success of query expansion lies in identifying appropriate expansion terms.
	
	In the literature, much effort has been devoted to investigating selection of proper additional terms. Among various studies, one promising line of research focuses on the pseudo-relevance feedback (PRF) \cite{DBLP:conf/sigir/XuC96,DBLP:journals/tois/XuC00,DBLP:conf/sigir/CaoNGR08,DBLP:conf/ecir/AlmasriBC16}, which assumes that the top-ranked documents returned from the first round retrieval are relevant for a given user query. The expansion terms are then extracted from the set of pseudo-relevance documents based on different term weighting schemes and selection criteria. The advantage of PRF is that the extracted terms are automatically obtained based on the context analysis of local document collections, and thus significantly bridge the gap between query and document vocabularies.
	
	Most of the aforementioned approaches are extractive, that is, the expansion terms are extracted from local relevance documents. In recent years, large-scale generative pre-trained language models such as GPT-2 \cite{radford2019language} and BART \cite{DBLP:conf/acl/LewisLGGMLSZ20} have achieved remarkable results on a variety of natural language processing tasks. However, despite the success of these neural text generation models, their performance in the query expansion scenario has been less studied.~\author{DBLP:journals/corr/abs-2009-08553}~\cite{DBLP:journals/corr/abs-2009-08553} proposed the Generation-Augmented Retrieval approach but it is specific to the problem of open-domain question answering.~\author{claveau2020query}~\cite{claveau2020query} presented to generate new query terms only with the GPT-2 model but without employing the rich local document information.~\author{DBLP:conf/kdd/LeeGZ18}~\cite{DBLP:conf/kdd/LeeGZ18} proposed the QE-CGAN model for rare query expansion but without using the pseudo-relevance feedback.
	
	To combine the merits of both the traditional pseudo-relevance feedback and recently developed powerful natural language generation models, in this article, we propose Generative Query Expansion with Pseudo-Relevance Feedback (GQE-PRF) approach. Our GQE-PRF model explores query expansion with a similar two-stage process as in the conventional PRF-based framework. First, given the initial query, we obtain the top-ranked documents through standard efficient sparse retrieval techniques such as BM25~\cite{DBLP:journals/ftir/RobertsonZ09} algorithm. Next, instead of extracting related terms from pseudo-relevance documents, we generate new expansion terms via neural text generation models. It is worth to emphasize that both the retrieved local documents and the original query are taken as inputs to the generative model. In this way, we can benefit from the rich local document information while still embracing the strong power of neural text generation.
	
	In addition, selection of  high quality terms is essential to the performance of query expansion. However, it has been known that top-ranked documents from initial retrieval may be irrelevant to the original query and thus can have potential negative effects. To alleviate this problem and better direct the query generation process, we adopt the Conditional Generative Adversarial Networks (CGANs) \cite{DBLP:journals/corr/MirzaO14} for training, and propose the novel PRF-CGAN framework which constructively integrates pseudo-relevance feedback into CGAN. To be more specific, the model contains a generator which is the same as the one in the GQE-PRF model, and a discriminator that distinguishes the generated query terms from the raw ones. Both the generator and the discriminator are conditioned on the pseudo-relevance documents and trained simultaneously. 
	
	We summarize the key contributions of this work as follows:
	\begin{itemize}
		\item We describe a new strategy to combine the powerful natural language generation models with the pseudo-relevance feedback based query expansion and propose the novel GQE-PRF approach.
		\item We present the PRF-CGAN method to train the generative query expansion model which extends the classical conditional generative adversarial networks with pseudo-relevance feedback.
		\item We demonstrate the effectiveness of our model on information retrieval tasks on benchmark datasets.
	\end{itemize}
	
	\section{Methodology}
	
	In this section, we introduce the details of our proposed methods. We first describe the GQE-PRF model, and then present the PRF-CGAN approach.
	
	\subsection{GQE-PRF Model}
	
	\begin{figure}
		\centering
		\includegraphics[width=0.9\linewidth]{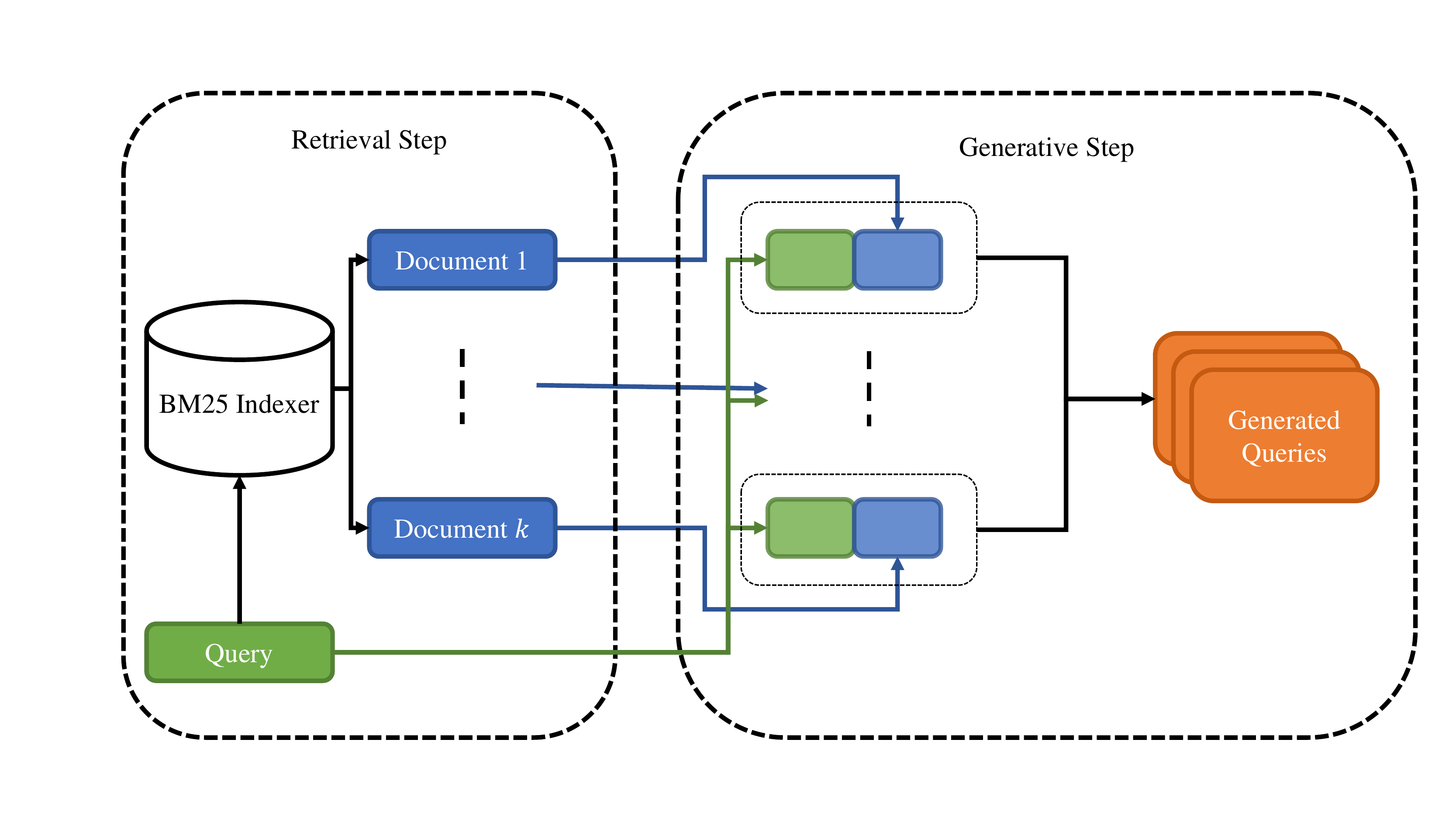}
		\caption{The overall architecture of the GQE-PRF model.}
		\label{fig:GQE-PRF}
	\end{figure}
	
	\subsubsection{Model Overview}
	Given an initial user query, we are aiming at generating related expansion terms with the benefits of both local relevant documents and pre-trained natural language generation models. The overall model framework is depicted in Figure~\ref{fig:GQE-PRF}. Our GQE-PRF model is in a retrieval-generation manner. Specifically, it mainly consists of two steps: (1) a pseudo-relevance feedback retrieval module and (2) an expansion term generation step. The details of each stage are explained in the following sections.
	
	\subsubsection {Initial Retrieval}
	The primary goal of the initial retrieval step in query expansion is to efficiently return the pseudo-relevance feedback. In general, classical term-based retrieval approaches, including TF-IDF and BM25, have proven to be sufficient for this purpose. Therefore, in our GQE-PRF model, we adopt BM25 method as our initial document retriever as used in various traditional PRF-based query expansion models. To be more specific, given an initial query $q$ and a document collection $\mathcal{D}$, BM25 model retrieves the top-$k$ document set $\mathcal{D}_k^q$  for the query $q$ from $\mathcal{D}$, which is then passed as inputs to the next generative step.
	
	\subsubsection {Expansion Term Generation}
	Recent progress in large-scale Pre-trained Language Models (PLMs) has considerably improved the performance of text generation. These PLMs are usually trained on large text corpora attempting to learn a general language representation, and then fine-tuned on specific downstream tasks. Several pre-trained text generation models, e.g., GPT-2 \cite{radford2019language} and BART \cite{DBLP:conf/acl/LewisLGGMLSZ20}, have been proposed and studied extensively. These neural generation models mostly adopt a common encoder-decoder sequence-to-sequence architecture. In particular, the encoder takes the initial sequence as input and transforms it into an intermediate hidden representation, and then the decoder takes the hidden representations as input and generates the final output sequence.
	
	Since the main objective of query expansion is to explore proper related terms, it is natural to introduce text generation models to query expansion and thus identify relevant terms in a generative way. In this work, we focus on the BART model due to its wide range of applications in natural language generation tasks, but other neural generation models can also be implemented in general.
	
	Specifically, following the standard practice, we concatenate the initial query $q$ with each retrieved relevant document, separated by the special token \texttt{[SEP]}, as the input to the BART model, and then outputs the generated query terms.
	
	\subsection{PRF-CGAN Model}
	In order to train the expansion term generation model in GQE-PRF, in this section, we propose an associated approach which relies on the conditional generative adversarial networks. We first give a brief introduction to CGAN, and then present our PRF-CGAN framework and illustrate each component in details.
	
	\begin{figure}
		\centering
		\includegraphics[width=0.9\linewidth]{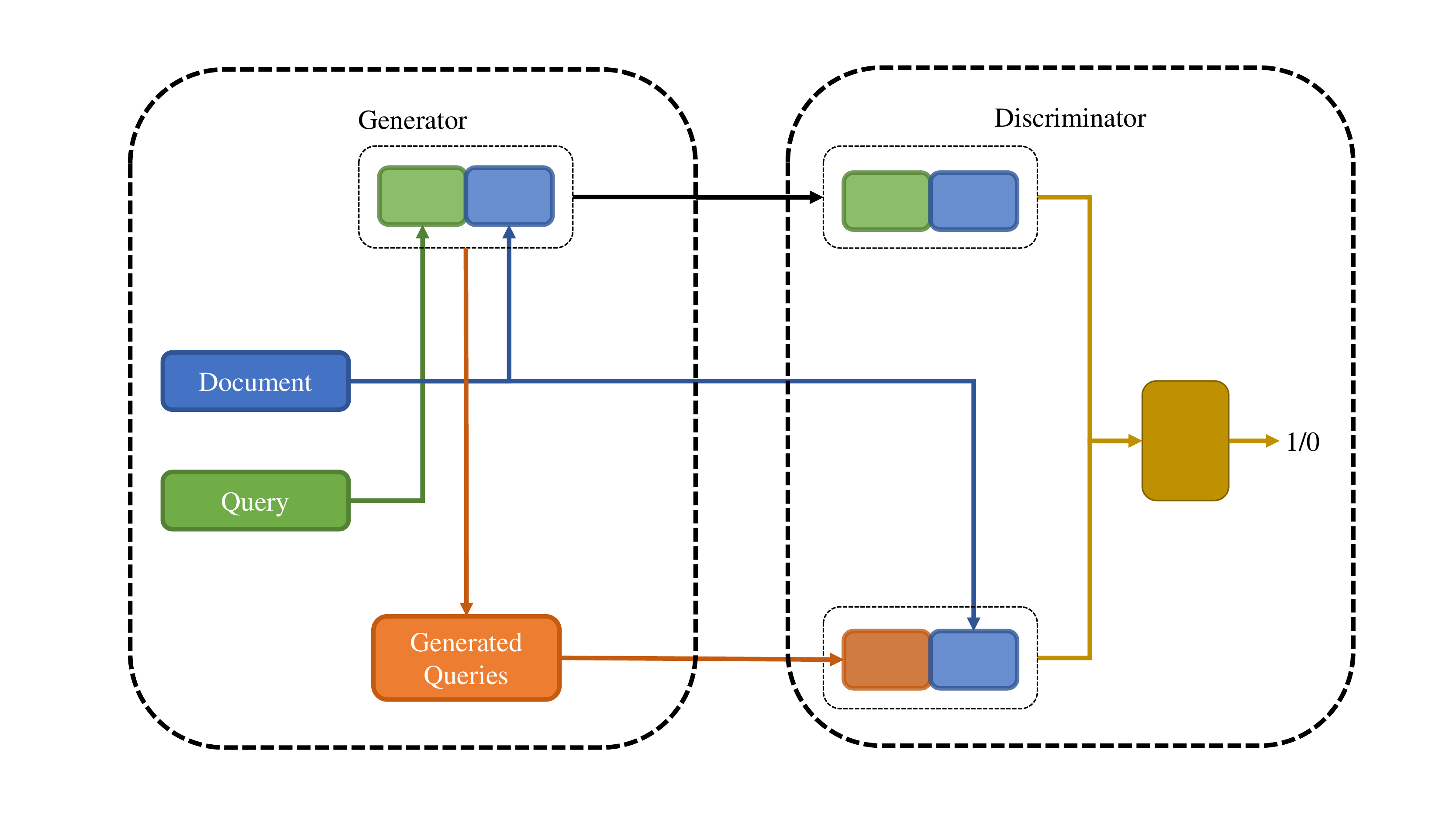}
		\caption{The framework of the PRF-CGAN approach.}
		\label{fig:PRF-CGAN}
	\end{figure}
	
	\subsubsection{CGAN Model}
	The conditional generative adversarial network is an extension of the standard Generative Adversarial Network (GAN)~\cite{DBLP:journals/corr/GoodfellowPMXWOCB14} framework. As in the GAN model, CGAN is also composed of two modules: a generator and a discriminator. The generator is used to produce new realistic samples, while the discriminator is trained to identify the generated ones from the true training data. Both the generator and the discriminator are optimized simultaneously based on an adversarial min-max problem. However, in contrast to the traditional GAN approach, both components in CGAN are conditioned on some extra auxiliary information, which in turn ensures better generation direction. For more details regrading the CGAN approach, please refer to \cite{DBLP:journals/corr/MirzaO14}.
	
	\subsubsection{PRF-CGAN Framework}
	We propose the PRF-based CGAN (PRF-CGAN) framework to train the GQE-PRF model. The key objective of PRF-CGAN is to explore a better training strategy for a more controllable generation process. Figure~\ref{fig:PRF-CGAN} describes the overall model architecture. In addition to the general framework of GAN, our model takes the pseudo-revelance feedback as inputs for both the generator and the discriminator and allows the generation process towards the direction of rich information contained in the local documents.
	
	\subsubsection{Generator}
	The generator in the PRF-CGAN model is the same as the expansion term generation model in GQE-PRF. We adopt the BART model which takes both the initial query and pseudo-relevance documents as input and outputs the generated query terms.
	
	\subsubsection{Discriminator}
	Motivated by the CGAN framework, our discriminator is also conditioned on the pseudo-relevance feedback. More specifically, for a given query (either the initial query or the expanded one), we first concatenate it with the pseduo-relevance document separated by the special token \texttt{[SEP]}, and then pass the new sequence to the BART model, followed by obtaining the hidden representation of the last token of the input sequence. Next, we take the last hidden vector as input to a simple binary classifier to identify whether the query is from the true training data or from the generator. Moreover, we consider the sequences (query \texttt{[SEP]} relevant document) as positive training samples for the binary classifier, and the concatenations (query \texttt{[SEP]} irrelevant document) or the sequences (expanded query \texttt{[SEP]} pseudo-relevance feedback) as negative samples.
	
	\section{Experiments And Results}
	\label{sec:expeiment}
	In this section, we validate the performance of our proposed GQE-PRF approach. Since the goal of query expansion is to improve the  quality of the follow-up retrieval and reranking systems, we examine the effectiveness of GQE-PRF in terms of the performance on the retrieval and reranking tasks. We first describe the exmperimental setup in Section~\ref{sec:expsetup}, then report the results on retrieval and reranking tasks in Section~\ref{sec:resretrieval} and ~\ref{sec:resrerank} respectively, and show the effect of hyperparameters in Section~\ref{sec:respara}. 
	
	\subsection{Experimental Setup}
	\label{sec:expsetup}
	\subsubsection{Datasets}
		\begin{table}
		\caption{Dataset statistics of ANTIQUE and wikIR1k. Columns correspond to number of queries, number of documents, average number of words per query, average number of words per document, and average number of relevant documents per query.}
		\label{tab:stats}
		\centering
		\begin{tabular}{|c|c|c|c|c|c|}
			\hline
			& \#q &\#d &\#words/q&\#words/d&\#rel. d/q\\
			\hline
			ANTIQUE&2,626&34,011&10.72&49.31&47.59\\
			\hline
			wikIR1k&1,644&370k&2.18&197.70&15.02\\	
			\hline
		\end{tabular}
	\end{table}
	We conduct extensive experiments on two public benchmark datasets to validate the effectiveness of our proposed approach: ANTIQUE~\cite{DBLP:conf/ecir/HashemiAZC20} and wikIR1k~\cite{DBLP:conf/lrec/FrejSC20}. The ANTIQUE dataset is a non-factoid question answering dataset. It consists of $34,011$ question-answer pairs with a four-level relevance annotations, including $2,426$ questions for training and $200$ for test. The wikIR1k dataset is an annotated information retrieval dataset built upon Wikipedia, containing $370$k documents, $1,444$, $100$ and $100$ queries for training, validation and test, respectively. The summary statistics regarding the characteristics of queries and documents of the two datasets are provided in Table~\ref{tab:stats}. On average, the ANTIQUE dataset has longer queries than wikIR1k, reflecting the fact that queries for question answering task are expressed more closer to natural language.

	\subsubsection{Baseline Models}
	We compare our proposed GQE-PRF framework with two common baseline approaches for query expansion: RM3~\cite{DBLP:conf/sigir/LavrenkoC01} and PRF~\cite{UCAM-CL-TR-356}. We use the implementations available in the Anserini~\cite{DBLP:conf/ecir/LinCTCCFIMV16} package\footnote{https://github.com/castorini/anserini} with default parameter settings for both the baseline approaches. 
	
	\subsubsection{Evaluation Metrics} For retrieval tasks, we use top-$k$ retrieval accuracy to measure the performance of each approach, which is defined as the fraction of top-$k$ retrieved documents that are related to the query. For reranking tasks, we adopt the Mean Average Precision (MAP) and the normalized Discounted Cumulative Gain (nDCG) evaluation metrics.
	
	\subsubsection{Implementation Details}
	Given an initial query, we first obtain the top-$10$ pseudo-relevance documents from the collection via the efficient BM25 algorithm provided by the Anserini package. Then we join the query with each returned document separated by the \texttt{[SEP]} token  and pass it to the BART model to generate related expansion terms. We adopt the HuggingFace \cite{wolf-etal-2020-transformers} PyTorch implementation of BART model\footnote{https://github.com/huggingface/transformers}, and truncate long sequence to $1,024$ tokens.
	
	\subsection{Results on Retrieval Tasks}
	\label{sec:resretrieval}
		\begin{table}[h]
		\caption{Results on retrieval tasks in terms of top-$k$ retrieval accuracy on the two datasets.}
		\label{tab:retrieval}
		\centering
		\begin{tabular}{|c|c|c|c|c|}
			\hline
			&\multicolumn{2}{|c|}{ANTIQUE}&\multicolumn{2}{|c|}{wikIR1k}\\
			\hline
			QE method&Top-5&Top-10&Top-5&Top-10\\
			\hline
			None&0.1130&0.1941&0.1531&0.1981\\
			\hline
			RM3&0.1105&0.1933&\bf 0.1550&0.2014\\
			\hline
			PRF&0.1065&0.1895&0.1535&\bf 0.2135\\
			\hline
			GQE-PRF&\bf 0.1132&\bf 0.1943&0.1540&0.2003\\	
			
			\hline
		\end{tabular}
	\end{table}
	To compare the performance on retrieval tasks with respect to different query expansion methods, we use BM25 model as our base retriever. For all experiments regarding BM25, we adopt the default implementations in the Anserini package. For the GQE-PRF approach, we formulate the new query by appending the top-$n$ generated expansion terms to the initial query, separated by space characters. Here, $n$ is a hyperparameter denoting the number of additional query terms and is tuned via a grid search. In this work, we choose $n$ to be $7$ by searching from the set $\{1,2,\ldots,10\}$.
	
	Table~\ref{tab:retrieval} reports the results on retrieval tasks in terms of top-$k$ retrieval accuracy. We compare GQE-PRF with three approaches: (1) the raw BM25 algorithm without any query expansion and (2) two baseline query expansion methods combined with BM25 retriever. It can be seen that our approach achieves improved results on the ANTIQUE dataset and comparable results on the wikIR1k dataset, showing that the GQE-PRF approach is robust to both long and short queries, while RM3 and PRF work better on keyword-based queries.	
	
	
	\subsection{Results on Reranking Tasks}
	\label{sec:resrerank}
		\begin{table}[h]
		\caption{Results on reranking tasks in terms of MAP and nDCG  measures on the two datasets.}
		\label{tab:reranking}
		\centering
		\begin{tabular}{|c|c|c|c|c|c|}
			\hline
			&&\multicolumn{2}{|c|}{ANTIQUE}&\multicolumn{2}{|c|}{wikIR1k}\\
			\hline
			Ranker&QE method&MAP&nDCG&MAP&nDCG\\
			\hline
			Vanilla BERT&None&0.3056&0.4628&0.2471&0.4818 \\
			\hline
			Vanilla BERT&RM3&0.1337&0.2783&0.1708&0.3500 \\
			\hline
			Vanilla BERT&PRF&0.1117&0.2493&0.1630&0.3569 \\		
			\hline
			Vanilla BERT&GQE-PRF&\bf 0.3082&\bf 0.4629&\bf 0.2493&\bf 0.4831 \\	
			\hline			
			CEDR-KNRM&None&0.3133&0.4615&0.2305&0.4637 \\
			\hline
			CEDR-KNRM&RM3&0.1584&0.3127&0.1613&0.3340 \\
			\hline
			CEDR-KNRM&PRF&0.1114&0.2522&0.1540&0.3357 \\	
			\hline	
			CEDR-KNRM&GQE-PRF&\bf 0.3146&\bf 0.4632&\bf 0.2366&\bf 0.4665 \\			
			\hline
		\end{tabular}
	\end{table}
	After the efficient first-step retrieval of relevant documents, the reranking stage aims at refining the candidate list through more sophisticated matching approaches. In the past few years, due to the powerful natural language understanding ability of BERT~\cite{DBLP:conf/naacl/DevlinCLT19} model, neural ranking models based on BERT have received considerable attention and achieved outstanding results. Accordingly, in this work, we assess the performance of query expansion for reranking tasks conditioned on BERT-based neural rankers.
	
	More specifically, we compare the reranking performance of GQE-PRF method with baseline query expansion approaches on two neural rankers, namely the Vanilla BERT model and the CEDR-KNRM~\cite{DBLP:conf/sigir/MacAvaneyYCG19} model. We adopt the implementations in the OpenNIR~\cite{DBLP:conf/wsdm/MacAvaney20} package\footnote{https://github.com/Georgetown-IR-Lab/OpenNIR} for the two rankers with reranking $100$, $5$ and $50$ top retrieved documents for training, validation and test, respectively.
	
	Table~\ref{tab:reranking} summarizes the results on reranking tasks. Our GQE-PRF approach consistently outperforms other baseline query expansion approaches on the two neural rankers and on the two datasets. This is mainly because the query terms generated by GQE-PRF approach are from the BART model and thus have more semantic meanings, which in turn enable the BERT-based rankers to better understand the intent of the original query and improve ranking performance.

	\subsection{Effect on the Number of Expansion Terms}
	\label{sec:respara}
		\begin{figure}
		\centering
		\includegraphics[width=0.8\linewidth]{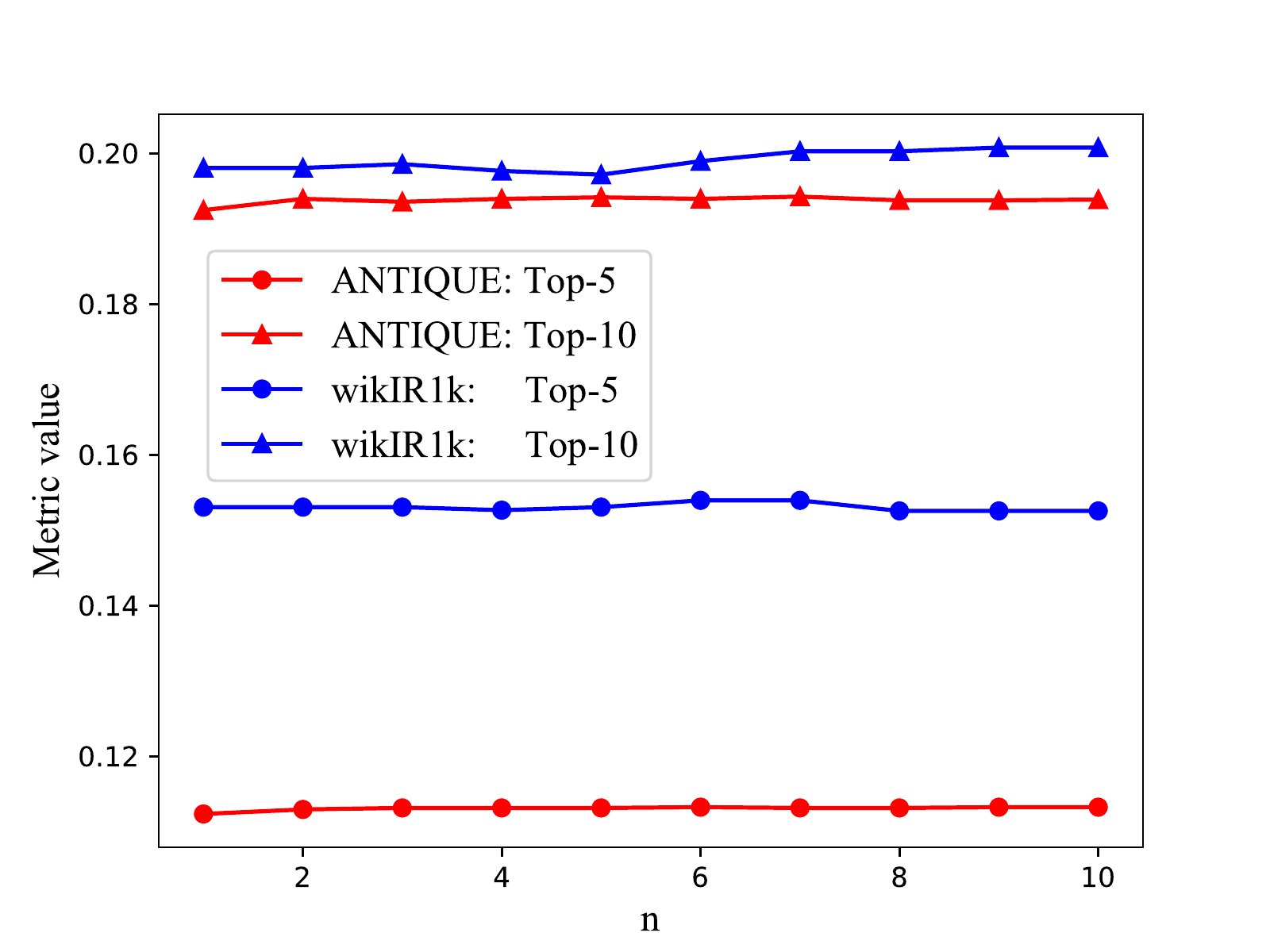}
		\caption{The top-$k$ retrieval accuracy of GQE-PRF combined with BM25 algorithm with different number of expansion terms $n$.}
		\label{fig:retrieval_with_k}
	\end{figure}
\begin{figure}
		\centering
		\includegraphics[width=0.8\linewidth]{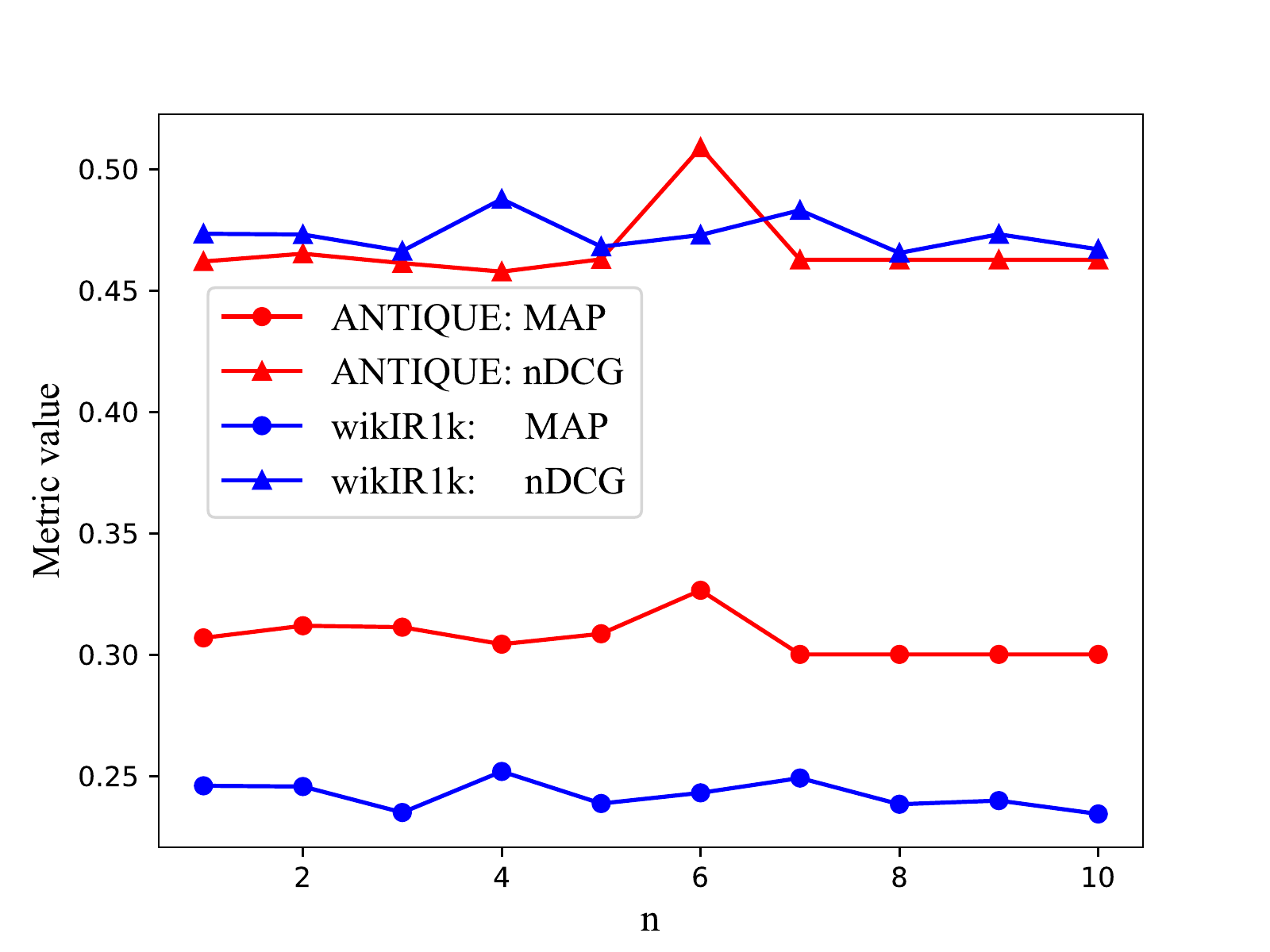}
		\caption{The MAP and nDCG results of Vanilla BERT+GQE-PRF with varying number of expansion terms $n$.}
		\label{fig:reranking_with_k}
	\end{figure}

	In this section, we study the effect on the number of query expansion terms $n$, which is a hyperparameter for the GQE-PRF approach. Figure~\ref{fig:retrieval_with_k} and Figure~\ref{fig:reranking_with_k} show the results with varying $n$ on the two datasets for the retrieval and reranking tasks, respectively. For the neural ranking models, they achieve the best results when $n=6$ on the ANTIQUE dataset, while $n=4$ for wikIR1k. However, for the retrieval tasks, the results exhibit little change with varying number of expansion terms. This indicates that BERT-based models are more sensitive to the semantic meaning of the queries.
	
	\section{Conclusion}
	We investigate the performance of large-scale pre-trained neural text generation models in the field of query expansion, and propose the GQE-PRF approach. Our experimental results show that integrating natural language generation models into PRF-based query expansion methods can improve the performance over the traditional term-based expansion approaches. Moreover, when combined with BERT-based neural ranking systems, our model consistently outperforms the baseline approaches, indicating that BART-based query expansion method can help the BERT-based rankers have a better understanding of the user intents.
	
	%
	%
	\bibliographystyle{splncs04}
	\bibliography{samplepaper}
	
\end{document}